\documentclass[11pt,fleqn]{article}

\usepackage[T1]{fontenc}
\usepackage[utf8]{inputenc}
\usepackage{lmodern}
\usepackage{amsmath}
\usepackage{amssymb}
\usepackage{mathtools}
\usepackage{bm}
\usepackage[top=30mm,bottom=30mm,left=25mm,right=25mm]{geometry}
\usepackage{xcolor}
\usepackage{hyperref}

\numberwithin{equation}{section}
\allowdisplaybreaks[3]
\hypersetup{
  unicode,
  colorlinks,
  citecolor=blue,
  linkcolor=blue,
  urlcolor=red,
  pdftitle={A Split-Signature Type IIB Background on AdS5,C},
  pdfauthor={Brenno Carlini Vallilo},
  bookmarksopen=true,
  bookmarksopenlevel=2
}

\newcommand{\ii}{\mathrm i}
\newcommand{\AdS}{\mathrm{AdS}}
\newcommand{\gC}{\mathbb C}
\newcommand{\gR}{\mathbb R}
\newcommand{\psl}{\mathfrak{psl}}
\newcommand{\slalg}{\mathfrak{sl}}
\newcommand{\spalg}{\mathfrak{sp}}
\newcommand{\soalg}{\mathfrak{so}}
\newcommand{\vol}{\operatorname{vol}}
\newcommand{\Ric}{\operatorname{Ric}}

\newcommand\blfootnote[1]{%
  \begingroup
  \renewcommand\thefootnote{}\footnote{#1}%
  \addtocounter{footnote}{-1}%
  \endgroup
}

\begin{document}

\vspace{36pt}
\begin{center}
{\LARGE A Split-Signature Type IIB Background on $\AdS_{5,\gC}$}

\vspace{30pt}
Brenno Carlini Vallilo${}^{\spadesuit}$

\vspace{25pt}
${}^{\spadesuit}${\em Departamento de F\'isica y Astronom\'ia,
Facultad de Ciencias Exactas,\\
Universidad Andres Bello,
Sazi\'e 2212, Santiago, Chile}

\vspace{40pt}
{\bf Abstract}
\end{center}
We construct a maximally supersymmetric Type IIB background in signature
$(5,5)$ on the real ten-manifold underlying complex $\AdS_5$.  The spacetime
is diffeomorphic to $T^*S^5$ and carries the invariant metric
$G=\ii(\bar q-q)$, where $q$ is the holomorphic constant-curvature metric.
The real five-form $F_5=(4\ii/L)(\Omega-\bar\Omega)$, built from its
normalized volume form, is self-dual, and its stress tensor supports this
scalar-flat, non-Einstein metric.  We derive these fields from the
most general $\mathrm{SL}(4,\gC)$-invariant metric and five-form ansatz:
self-duality and the Einstein equation fix their phases up to the overall
scale, orientation, and flux sign.  The fields obey the reality conditions
of the untwisted $\mathrm{IIB}_5$ theory.  The holomorphic coframe
makes the supercovariant curvature cancellation explicit, and flat
transport gives thirty-two global real Killing spinors.  Their Killing
superalgebra is the factor-exchanging real form of $\psl(4|4,\gC)$, and we
identify the corresponding supercoset.

\blfootnote{${}^{\spadesuit}$
\href{mailto:vallilo@unab.cl}{vallilo@unab.cl}}

\setcounter{page}{0}
\thispagestyle{empty}

\newpage
\tableofcontents
\newpage

\section{Introduction}

The $\AdS_5\times S^5$ solution occupies a central place in string theory
because it provides the original and best understood example of the AdS/CFT
correspondence: Type IIB string theory on this background is conjectured to be dual to
four-dimensional $\mathcal N=4$ super-Yang--Mills theory
\cite{Maldacena1998LargeN,GubserKlebanovPolyakov1998GaugeCorrelators,
Witten1998AdSHolography}.  The metric is the product of the constant-curvature
metrics on $\AdS_5$ and $S^5$, supported by a self-dual Ramond--Ramond
five-form, and the Killing superalgebra is $\mathfrak{psu}(2,2|4)$.
Once the solution is complexified, however, these
data do not select a unique real spacetime.  An antilinear involution may
preserve the two five-dimensional factors separately, as in the usual
Lorentzian background, or it may exchange them.  In this work we ask whether
the fixed locus of the factor-exchanging involution is itself a real Type IIB
solution.

The fixed locus is the underlying real ten-manifold of the complex quadric
\begin{equation}
  X=\AdS_{5,\gC}
  \simeq\frac{\mathrm{SL}(4,\gC)}{\mathrm{Sp}(4,\gC)}.
  \label{eq:intro-complex-quadric}
\end{equation}
Let $q$ denote its invariant holomorphic metric of curvature $-L^{-2}$,
and let $\Omega$ be the corresponding normalized holomorphic volume form.
We find a solution of Type IIB supergravity with constant axion and
dilaton, vanishing three-form fields, and metric and self-dual five-form
given by
\begin{equation}
  G=\ii(\bar q-q),
  \qquad
  F_5=\frac{4\ii}{L}(\Omega-\bar\Omega).
  \label{eq:intro-background}
\end{equation}
The metric has signature $(5,5)$ and is anti-Hermitian with respect to the
complex structure.  Its Ricci tensor is proportional to $q+\bar q$, so it
is not proportional to $G$, although its scalar curvature vanishes.  The
five-form stress tensor supports this metric.  Self-duality and the
Einstein equation determine the phases in the general homogeneous ansatz,
leaving the scale, orientation, and sign of the flux.

Type IIB theories in signature $(5,5)$ admit Majorana--Weyl spinors and real
self-dual five-forms
\cite{Hull1998Signature,GallMohaupt2021ArbitrarySignature}.  Real parts of
holomorphic metrics give anti-K\"ahler geometries, and the complex sphere
has long been used as an example
\cite{BorowiecEtAl1999AlmostComplex}.  Split-signature branes and product
near-horizon geometries were studied in
\cite{HullKhuri1999Worldvolume}.  More recently, complex Type IIB
supergravity and its real forms were organized systematically in
\cite{BergshoeffEtAl2007AlternateRealities,
DHokerGutperleUhlemann2025ComplexIIB}.  We use this real form to construct
the background above and determine its Killing superalgebra.

We compute the metric, frame, connection, and curvature directly on the
quadric and use them to solve the homogeneous field equations.  In the
holomorphic coframe, the spin
curvature cancels the five-form contribution to the supercovariant
curvature.  The compatible real structure and the topology of $T^*S^5$
then give thirty-two global real Killing spinors.  This establishes maximal
supersymmetry as well as the real Killing superalgebra.

This paper is organized as follows. Section~\ref{sec:exchange-geometry} develops the geometry and Hodge operator.
We determine the invariant Type IIB solution in
Section~\ref{sec:iib-solution}, prove maximal supersymmetry in
Section~\ref{sec:supersymmetry}, and identify the Killing superalgebra and
supercoset in Section~\ref{sec:superisometry}.
Section~\ref{sec:comparison} compares the construction with known
backgrounds and discusses the worldsheet problem.  The appendices give the
RR convention dictionary, a real Clifford representation, and the complete
indexed superalgebra.

\section{Geometry of the realified quadric}
\label{sec:exchange-geometry}
\label{sec:split-geometry}

\subsection{The complex quadric as a homogeneous space}

We first describe the bosonic space in a form that makes both its curvature
and its realification explicit.  We use the affine quadric
\begin{equation}
  X
  =
  \left\{
    Z\in\gC^6\,\middle|\,
    Z\cdot Z=-L^2
  \right\}
  \simeq
  \frac{\mathrm{SO}(6,\gC)}{\mathrm{SO}(5,\gC)}.
  \label{eq:complex-quadric}
\end{equation}
Over $\gC$, the sign in the defining equation contains no signature
information.  The ambient quadratic form induces a homogeneous holomorphic
metric, which we denote by $q$.  The parameter $L$ fixes its complex curvature
scale, with $L>0$.  The signature enters when we combine this metric with
its conjugate to obtain a real spacetime metric.

To later connect this geometry with the Type IIB superalgebra, we rewrite the
quadric using $\mathrm{SL}(4,\gC)$.  Let $V=\gC^4$ carry the symplectic form
$\omega_{AB}$.  The orbit map
\begin{equation}
  g\,\mathrm{Sp}(4,\gC)
  \longmapsto
  Y_{AB}=L(g\omega g^{\mathsf T})_{AB}
  \label{eq:pfaffian-orbit}
\end{equation}
identifies the coset with the hypersurface of two-forms with fixed
Pfaffian.  Indeed,
\begin{equation}
  \operatorname{Pf}(Y)
  =
  \frac18\epsilon^{ABCD}Y_{AB}Y_{CD},
  \label{eq:pfaffian}
\end{equation}
is a quadratic constraint on the six-dimensional space $\wedge^2V$.  The
low-rank isomorphisms
$\slalg(4,\gC)\simeq\soalg(6,\gC)$ and
$\spalg(4,\gC)\simeq\soalg(5,\gC)$ give
\begin{equation}
  X
  \simeq
  \frac{\mathrm{SL}(4,\gC)}{\mathrm{Sp}(4,\gC)}.
  \label{eq:sl4-sp4-coset}
\end{equation}

The tangent representation is obtained directly from the symmetric-pair
decomposition
\begin{equation}
  \slalg(4,\gC)
  =
  \spalg(4,\gC)\oplus\mathfrak m,
  \qquad
  \mathfrak m
  \simeq
  \left(\wedge^2\mathbf4\right)_0
  \simeq
  \mathbf5,
  \qquad
  [\mathfrak m,\mathfrak m]
  \subset
  \spalg(4,\gC).
  \label{eq:symmetric-pair}
\end{equation}
The isotropy representation $\mathbf5$ is irreducible over $\gC$.  It
follows that the invariant holomorphic metric and volume form are unique up
to normalization.  This observation will reduce the homogeneous Type IIB
equations to equations for an overall scale and two phases.

\subsection{The underlying real manifold}

The real Type IIB spacetime is the underlying real manifold $X_{\gR}$ of
$X$.  Its complexified tangent bundle has the canonical splitting
\begin{equation}
  TX_{\gR}\otimes_{\gR}\gC
  \simeq
  T^{1,0}X\oplus T^{0,1}X,
  \qquad
  \overline{T^{1,0}X}=T^{0,1}X.
  \label{eq:realification-tangent-splitting}
\end{equation}
Conjugation exchanges the two summands, so neither descends separately to
a real five-dimensional tangent distribution.  To determine the
ten-manifold globally, we write
$Z=x+\ii y$, with $x,y\in\gR^6$.  The quadric equation becomes
\begin{equation}
  x^2-y^2=-L^2,
  \qquad
  x\cdot y=0.
  \label{eq:real-imaginary-quadric}
\end{equation}
These two equations have a simple tangent-bundle parametrization.  For
$n\in S^5$ and $p\in T_nS^5$, we define
\begin{equation}
  Z(n,p)
  =
  p+\ii\sqrt{L^2+|p|^2}\,n.
  \label{eq:tangent-bundle-map}
\end{equation}
The inverse is $p=x$ and $n=y/|y|$, which is well defined since
$|y|^2=L^2+|x|^2>0$.  The constraint $x\cdot y=0$ gives $p\in T_nS^5$.
We therefore obtain the global identification
\begin{equation}
  X_{\gR}
  \simeq
  TS^5
  \simeq
  T^*S^5
  \label{eq:global-topology}
\end{equation}
as real manifolds, where the last isomorphism uses the round metric on
$S^5$.  The complex structure inherited from the quadric becomes the
adapted complex structure on the tangent bundle.

\subsection{The coordinate metric, frame, and curvature}

We now compute the tensors that enter the field equations.  On an affine
patch we write $Z=(z^1,\ldots,z^5,W)$ and solve the quadric constraint for
$W$.  With $z^2=\delta_{ij}z^i z^j$, we have
\begin{equation}
  W^2=-L^2-z^2.
  \label{eq:affine-constraint}
\end{equation}
Differentiating gives $dW=-z_i dz^i/W$.  Substitution into the ambient
metric $q=\delta_{ij}dz^i dz^j+dW^2$ yields
\begin{equation}
  q_{ij}
  =
  \delta_{ij}
  -
  \frac{z_i z_j}{L^2+z^2},
  \qquad
  q^{ij}
  =
  \delta^{ij}
  +
  \frac{z^i z^j}{L^2}.
  \label{eq:affine-metric}
\end{equation}
Indices $i,j$ label the five complex coordinates; ambient coordinate
indices are raised and lowered with $\delta_{ij}$.  Inserting this metric
and its inverse into the Levi--Civita formula gives
\begin{equation}
  \Gamma^i{}_{jk}
  =
  -\frac{z^i}{L^2}q_{jk}.
  \label{eq:affine-christoffel}
\end{equation}
We use
$R^i{}_{jk\ell}=\partial_k\Gamma^i{}_{\ell j}
-\partial_\ell\Gamma^i{}_{kj}
+\Gamma^i{}_{km}\Gamma^m{}_{\ell j}
-\Gamma^i{}_{\ell m}\Gamma^m{}_{kj}$.
Differentiating the connection then gives
\begin{equation}
  R^i{}_{j k\ell}(q)
  =
  -\frac1{L^2}
  \left(
    \delta^i_k q_{j\ell}
    -
    \delta^i_\ell q_{jk}
  \right),
  \qquad
  \Ric_{ij}(q)=-\frac4{L^2}q_{ij}.
  \label{eq:holomorphic-curvature}
\end{equation}
Thus the induced metric has constant holomorphic curvature $-L^{-2}$.

An explicit coframe makes the corresponding spin connection available.
Choose a local branch
\begin{equation}
  h=(L^2+z^2)^{1/2},\qquad W=\ii h,
  \label{eq:affine-frame-branch}
\end{equation}
with $h\neq0$ and $h+L\neq0$.  Frame indices $a,b=1,\ldots,5$ are contracted
with $\delta_{ab}$.  The coframe
\begin{equation}
  E^a=dz^a-\frac{z^a z_b\,dz^b}{h(h+L)},
  \qquad q=\delta_{ab}E^a E^b
  \label{eq:affine-coframe}
\end{equation}
reproduces the coordinate metric: the coefficient of $z_i z_j$ in
$\delta_{ab}E^a{}_i E^b{}_j$ reduces to $-h^{-2}$.
Since $dh=z_b\,dz^b/h$, direct differentiation gives
\begin{equation}
  dE^a=\frac{z_c\,dz^c\wedge dz^a}{h(h+L)},
  \qquad
  \omega^a{}_b=\frac{z_b dz^a-z^a dz_b}{L(h+L)},
  \qquad
  dE^a+\omega^a{}_b\wedge E^b=0.
  \label{eq:affine-spin-connection}
\end{equation}
The connection is antisymmetric, so the last equation identifies it as the
torsion-free metric connection.  Differentiating its coefficient gives
$d\bigl(1/[L(h+L)]\bigr)=-z_c\,dz^c/[Lh(h+L)^2]$.  We obtain
\begin{align}
  \mathcal R^{ab}
  &:=d\omega^{ab}+\omega^a{}_c\wedge\omega^{cb}
  \nonumber\\
  &=-\frac1{L^2}dz^a\wedge dz^b
  +\frac{z^b z_c\,dz^a\wedge dz^c+z^a z_c\,dz^c\wedge dz^b}
        {L^2h(h+L)}
  =-\frac1{L^2}E^a\wedge E^b.
  \label{eq:affine-cartan-curvature}
\end{align}
This is the frame form of the coordinate curvature above.

The determinant of the coframe is $L/h$.  It fixes the volume form and
reproduces the metric determinant:
\begin{equation}
  \det(q_{ij})=\frac{L^2}{h^2}=-\frac{L^2}{W^2},
  \qquad
  \Omega=E^1\wedge\cdots\wedge E^5
  =\frac{\ii L}{W}\,dz^1\wedge\cdots\wedge dz^5.
  \label{eq:affine-volume}
\end{equation}
The $\mathrm{SO}(5,\gC)$ connection preserves this volume form.  Its
parallelism, closure, and normalization are
\begin{equation}
  \nabla_i\Omega_{jklmn}=0,\qquad d\Omega=0,\qquad
  \frac1{5!}\Omega_{i_1\ldots i_5}\Omega^{i_1\ldots i_5}=1,
  \label{eq:omega-normalization}
\end{equation}
where $\nabla_i$ is the Levi--Civita derivative of $q$ and indices in the
last contraction are raised with $q^{-1}$.  Since the connection is
torsion-free, the second equation follows from the first.
The coordinate formulas are local expressions for the invariant metric
and volume form on the quadric.

\subsection{Invariant real metrics and their connection}

We next construct the invariant metrics on $X_{\gR}$.  Complexifying its
real isotropy representation gives a holomorphic $\mathbf5$ and an
antiholomorphic $\mathbf5$, acted on by the two independent complexified
isotropy factors.  Each admits one invariant symmetric bilinear form,
while their tensor product contains no scalar.  The invariant real
metrics therefore have only $(2,0)+(0,2)$ components and take the form
\begin{equation}
  G_{\rho,\theta}
  =
  \rho
  \left(
    e^{-\ii\theta}q
    +
    e^{\ii\theta}\bar q
  \right),
  \qquad
  \rho>0.
  \label{eq:phase-metric}
\end{equation}
The real and imaginary parts of $E^a$ form a real coframe.  In this
coframe, the metric has the matrix
\begin{equation}
  [G_{\rho,\theta}]_{(\operatorname{Re}E,\operatorname{Im}E)}
  =2\rho\begin{pmatrix}
    \cos\theta\,\mathbf1_5&\sin\theta\,\mathbf1_5\\
    \sin\theta\,\mathbf1_5&-\cos\theta\,\mathbf1_5
  \end{pmatrix}.
  \label{eq:phase-real-metric}
\end{equation}
Its eigenvalues are $+2\rho$ and $-2\rho$, each with multiplicity five.
Thus every member is anti-Hermitian, $G(Ju,Jv)=-G(u,v)$, and has
\begin{equation}
  \operatorname{sign}(G_{\rho,\theta})=(5,5).
  \label{eq:split-signature}
\end{equation}
On the complexified tangent bundle the metric is block diagonal,
\begin{equation}
  G_{\rho,\theta}^{\gC}
  =
  \rho e^{-\ii\theta}q
  \oplus
  \rho e^{\ii\theta}\bar q.
  \label{eq:complex-metric-blocks}
\end{equation}
The constant factors cancel between the metric and its inverse in the
Christoffel symbols.  Holomorphicity also makes all mixed connection
coefficients vanish.  The Levi--Civita connection is therefore
$\nabla^q\oplus\overline{\nabla^q}$ for every $\rho,\theta$.
Its Ricci tensor is
\begin{equation}
  \Ric_{MN}(G_{\rho,\theta})
  =
  -\frac4{L^2}(q+\bar q)_{MN}.
  \label{eq:phase-ricci}
\end{equation}
Here $M,N=1,\ldots,10$ denote real tangent indices.  In complex coordinates,
\begin{equation}
  \Ric_{ij}(G_{\rho,\theta})
  =
  -\frac4{L^2}q_{ij},
  \qquad
  \Ric_{\bar i\bar j}(G_{\rho,\theta})
  =
  -\frac4{L^2}\bar q_{\bar i\bar j},
  \qquad
  \Ric_{i\bar j}(G_{\rho,\theta})=0.
  \label{eq:phase-ricci-blocks}
\end{equation}
The covariant Ricci tensor is independent of the phase and scale.  Taking
its trace introduces the inverse metric and gives
\begin{equation}
  R(G_{\rho,\theta})=-\frac{40}{\rho L^2}\cos\theta.
  \label{eq:phase-scalar-curvature}
\end{equation}

For the phase $\theta=\pi/2$ appearing in the background, set
\begin{equation}
  G_-
  :=
  G_{1,\pi/2}
  =
  \ii(\bar q-q).
  \label{eq:g-minus}
\end{equation}
The real coframe
\begin{equation}
  e^a=\operatorname{Re}E^a+\operatorname{Im}E^a,\qquad
  f^a=\operatorname{Re}E^a-\operatorname{Im}E^a,\qquad
  G_-=\sum_a\bigl((e^a)^2-(f^a)^2\bigr)
  \label{eq:real-orthonormal-frame}
\end{equation}
makes the split signature explicit throughout the patch.  Writing
$\omega=A+\ii B$, the torsion equation gives the real connection
\begin{equation}
  d\begin{pmatrix}e\\f\end{pmatrix}
  +\varpi\wedge\begin{pmatrix}e\\f\end{pmatrix}=0,
  \qquad
  \varpi=\begin{pmatrix}A&B\\-B&A\end{pmatrix}.
  \label{eq:real-spin-connection}
\end{equation}
Both $A$ and $B$ are antisymmetric, so $\varpi$ takes values in
$\soalg(5,5)$.  Its curvature is the same real representation of
$\mathcal R=d\omega+\omega\wedge\omega$:
\begin{equation}
  d\varpi+\varpi\wedge\varpi
  =\begin{pmatrix}
    \operatorname{Re}\mathcal R&\operatorname{Im}\mathcal R\\
    -\operatorname{Im}\mathcal R&\operatorname{Re}\mathcal R
  \end{pmatrix}.
  \label{eq:real-cartan-curvature}
\end{equation}
For this metric, the Ricci tensor and scalar curvature are
\begin{equation}
  \Ric_{MN}(G_-)
  =
  -\frac4{L^2}(q+\bar q)_{MN},
  \qquad
  R(G_-)=0.
  \label{eq:crossed-ricci}
\end{equation}
Since $q+\bar q$ and $\ii(\bar q-q)$ are not proportional, the metric is
not Einstein.  The five-form must reproduce the nonzero tensor
$-4L^{-2}(q+\bar q)$ while leaving the scalar curvature zero.

\subsection{The middle-dimensional Hodge operator}

To impose five-form self-duality, choose the orientation for which the
unit-scale metric volume is
\begin{equation}
  \vol_{10}
  =
  \ii\Omega\wedge\bar\Omega,
  \qquad
  \vol(G_{\rho,\theta})=\rho^5\vol_{10}.
  \label{eq:orientation}
\end{equation}
On real five-forms in signature $(5,5)$, the Hodge operator obeys
\begin{equation}
  *^2=(-1)^{5(10-5)+5}=1.
  \label{eq:star-square}
\end{equation}
The Hodge action for the full phase family can be obtained without choosing
real coordinates.  Each raised holomorphic index contributes
$\rho^{-1}e^{\ii\theta}$, whereas the metric volume contributes $\rho^5$.
The scale cancels because the forms have middle degree.  Using
$\alpha\wedge *\beta=\langle\alpha,\beta\rangle\vol$, we find
\begin{equation}
  *_{\rho,\theta}\Omega
  =
  \ii e^{5\ii\theta}\bar\Omega,
  \qquad
  *_{\rho,\theta}\bar\Omega
  =
  -\ii e^{-5\ii\theta}\Omega.
  \label{eq:phase-hodge-star}
\end{equation}
For the phase of $G_-$, this becomes
\begin{equation}
  *_{G_-}\Omega=-\bar\Omega,
  \qquad
  *_{G_-}\bar\Omega=-\Omega.
  \label{eq:g-minus-hodge-star}
\end{equation}
Thus the real combination $\ii(\Omega-\bar\Omega)$ is self-dual.

\section{The homogeneous \texorpdfstring{$\mathrm{IIB}_5$}{IIB5} solution}
\label{sec:iib-solution}

\subsection{Reality condition}

We now embed the invariant tensors into Type IIB supergravity.  We keep the
axion and dilaton constant.  Since the Einstein-frame metric and $F_5$ are
invariant under the global $\mathrm{SL}(2,\gR)$ symmetry, we choose the scalar
duality frame
\begin{equation}
  \chi=0,
  \qquad
  \phi=0,
  \label{eq:constant-scalar-frame}
\end{equation}
and set all form fields other than $F_5$ to zero.
In signature $(5,5)$, the untwisted spinor condition is
\begin{equation}
  \mathcal B^{-1}_{(5,5)}\epsilon^*=\epsilon.
  \label{eq:iib5-spinor-reality}
\end{equation}
Here $\epsilon$ is the Type IIB spinor doublet, and
$\mathcal B_{(5,5)}$ is the complex-conjugation matrix defined in
Appendix~\ref{app:clifford-check}; it acts on each spinor separately.
On this branch, the metric, axion, two-form potentials, and four-form
potential are real.  It is denoted $\mathrm{IIB}_5$ in
\cite{BergshoeffEtAl2007AlternateRealities,
DHokerGutperleUhlemann2025ComplexIIB}.  The twisted branches
$\mathrm{IIB}_5^\star$ and $\mathrm{IIB}_5^\prime$ instead assign an
imaginary reality condition to the four-form potential.  The flux
constructed below is real, so the realified background belongs to the
untwisted $\mathrm{IIB}_5$ theory.

\subsection{Field equations and normalization}

Our five-form conventions are
\begin{equation}
  F_5
  =
  \frac1{5!}F_{M_1\ldots M_5}
  dx^{M_1}\wedge\cdots\wedge dx^{M_5},
  \qquad
  F_5\mathbin{\cdot}F_5
  =
  \frac1{5!}F_{M_1\ldots M_5}F^{M_1\ldots M_5}.
  \label{eq:five-form-contraction-convention}
\end{equation}
The untwisted real form with five timelike directions retains the usual
sign of the five-form term
\cite{BergshoeffEtAl2007AlternateRealities}.  In the chosen scalar frame,
the metric--five-form pseudo-action is
\begin{equation}
  S_{\mathrm{ps}}
  =
  \frac1{2\kappa_{10}^2}
  \int d^{10}x\,\sqrt{|G|}
  \left(
    R
    -\frac1{4\cdot5!}
     F_{M_1\ldots M_5}F^{M_1\ldots M_5}
  \right),
  \qquad
  F_5=*F_5,
  \label{eq:iib5-pseudo-action}
\end{equation}
We vary the pseudo-action before imposing self-duality.  A self-dual
five-form has zero scalar contraction, since
\begin{equation}
  \frac1{5!}F_{PQRST}F^{PQRST}\vol(G)
  =
  F_5\wedge*F_5
  =
  F_5\wedge F_5
  =0.
  \label{eq:self-dual-five-form-null}
\end{equation}
The traced Einstein equation consequently gives $R=0$, and the equations
in this truncation are
\begin{equation}
  dF_5=0,
  \qquad
  F_5=*F_5,
  \qquad
  R_{MN}
  =
  \frac1{96}
  F_{MPQRS}F_N{}^{PQRS}.
  \label{eq:iib-truncated-equations}
\end{equation}
The corresponding gravitino operator is
\cite{BergshoeffEtAl2007AlternateRealities,
HackettJonesSmith2004IIBKilling}
\begin{equation}
  \mathcal D_M
  =
  \nabla_M
  +
  \frac1{16\cdot5!}
  F_{ABCDE}\Gamma^{ABCDE}\Gamma_M\mathcal J,
  \qquad
  \mathcal J=\ii\sigma_2,
  \qquad
  \mathcal J^2=-1.
  \label{eq:gravitino-operator}
\end{equation}
Here $\mathcal J$ acts on the Type IIB spinor doublet.
Appendix~\ref{app:rr-conventions} gives the
metric variation and the conversion to the alternative RR convention.

\subsection{Determination of the invariant phase}

We now solve the field equations for the invariant tensors.  Under the
complexified isotropy, the exterior powers of each vector $\mathbf5$
contain a scalar only in degrees zero and five.  Consequently the invariant
five-forms are spanned by $\Omega$ and $\bar\Omega$.  Together with the
metric family already determined, the general
$\mathrm{SL}(4,\gC)$-invariant ansatz is
\begin{equation}
  G=G_{\rho,\theta},\qquad
  F_5=a\Omega+\bar a\bar\Omega,
  \qquad \rho>0,\quad a\in\gC.
  \label{eq:general-invariant-ansatz}
\end{equation}
The coefficients are constant, so $dF_5=0$.  The scalar curvature first
provides a useful restriction: the traced Einstein equation and
\eqref{eq:phase-scalar-curvature} require $\cos\theta=0$.  The full
self-duality and Einstein equations determine which phase is compatible
with the chosen orientation and fix the flux coefficient.

Using the Hodge action on $\Omega$, self-duality gives
\begin{equation}
  \bar a
  =
  \ii e^{5\ii\theta}a.
  \label{eq:rigidity-self-duality}
\end{equation}
For the Einstein equation, the volume normalization implies
\begin{equation}
  \Omega_{i p q r s}
  \Omega_j{}^{p q r s}
  =4!\,q_{ij},
  \label{eq:volume-contraction}
\end{equation}
where the four contracted indices are raised with $q^{-1}$.  Raising
them instead with $G^{-1}$ contributes
$\rho^{-4}e^{4\ii\theta}$.  The holomorphic Einstein equation is therefore
\begin{equation}
  \frac14
  a^2\rho^{-4}e^{4\ii\theta}
  =
  -\frac4{L^2},
  \qquad\text{or}\qquad
  a^2e^{4\ii\theta}
  =
  -\frac{16\rho^4}{L^2}.
  \label{eq:rigidity-einstein}
\end{equation}
Solving this equation for $a$ gives
\begin{equation}
  a
  =
  s\frac{4\ii\rho^2}{L}e^{-2\ii\theta},
  \qquad
  s=\pm1.
  \label{eq:rigidity-a-solution}
\end{equation}
We then substitute this result into the self-duality condition.  All scales
and the sign $s$ cancel, leaving
\begin{equation}
  e^{\ii\theta}=\ii.
  \label{eq:rigidity-phase}
\end{equation}
For the orientation \eqref{eq:orientation}, the invariant solution is
therefore
\begin{equation}
  \theta=\frac\pi2\quad\bmod 2\pi,
  \qquad
  a=\pm\frac{4\ii\rho^2}{L}.
  \label{eq:rigidity-result}
\end{equation}
The remaining parameter $\rho$ acts by the standard supergravity scaling
$G\mapsto\rho G$ and $F_5\mapsto\rho^2F_5$.
Up to this scale, orientation, and $F_5\mapsto-F_5$, the solution is unique
within the $\mathrm{SL}(4,\gC)$-invariant ansatz
\eqref{eq:general-invariant-ansatz}.

\subsection{The metric and five-form}

Choosing $\rho=1$ and the positive flux sign gives
\begin{equation}
  G=G_-=\ii(\bar q-q),\qquad
  F_5=\frac{4\ii}{L}(\Omega-\bar\Omega).
  \label{eq:solution-five-form}
\end{equation}
The flux is real and closed.  The Hodge relation gives
\begin{equation}
  *_{G_-}F_5=F_5.
  \label{eq:solution-self-duality}
\end{equation}
The stress tensor can now be read off directly.  On the holomorphic
block the four inverse metrics contribute $\ii^4=1$, so
\begin{equation}
  \frac1{96}
  F_{iPQRS}F_j{}^{PQRS}
  =
  \frac1{96}
  \left(\frac{4\ii}{L}\right)^2
  4!\,q_{ij}
  =
  -\frac4{L^2}q_{ij}.
  \label{eq:holomorphic-einstein-check}
\end{equation}
The antiholomorphic block is its conjugate and the mixed block vanishes.
Thus
\begin{equation}
  \Ric_{MN}(G_-)
  =
  \frac1{96}F_{5\,MPQRS}F_{5\,N}{}^{PQRS}.
  \label{eq:einstein-solved}
\end{equation}
\section{Maximal supersymmetry}
\label{sec:supersymmetry}

The constant axion and dilaton and the vanishing three-form fields make
the dilatino variation vanish for every supersymmetry parameter
\cite{DHokerGutperleUhlemann2025ComplexIIB}.  We therefore need only solve
the gravitino equation $\mathcal D\epsilon=0$.  The five-form is parallel:
$\nabla^q\Omega=0$ and its conjugate imply $\nabla F_5=0$ for the real
Levi--Civita connection.

\subsection{The holomorphic coframe}

We use the coframe $E^a$ and its conjugate to write the complexified
background as
\begin{equation}
  G_-^{\gC}
  =
  -\ii\delta_{ab}E^a E^b
  +\ii\delta_{ab}\bar E^a\bar E^b,
  \qquad
  F_5^{\gC}=\frac{4\ii}{L}(\Omega-\bar\Omega).
  \label{eq:complex-frame-background}
\end{equation}
The holomorphic and antiholomorphic distributions are orthogonal, with
sectional curvatures $-\ii/L^2$ and $\ii/L^2$, respectively.  Each flux
component is proportional to the volume form of its distribution, giving
the complex Freund--Rubin structure directly in this coframe.

Let $E_a,\bar E_a$ be the dual vectors, and write
$\Gamma_a=\Gamma(E_a)$ and $\Gamma_{\bar a}=\Gamma(\bar E_a)$.
The metric phases enter their Clifford relations explicitly:
\begin{equation}
  \{\Gamma_a,\Gamma_b\}=-2\ii\delta_{ab},
  \qquad
  \{\Gamma_{\bar a},\Gamma_{\bar b}\}=2\ii\delta_{ab},
  \qquad
  \{\Gamma_a,\Gamma_{\bar b}\}=0.
  \label{eq:holomorphic-clifford-relations}
\end{equation}

\subsection{Supercovariant flatness}

We compute the curvature of the gravitino
operator \eqref{eq:gravitino-operator}.  In the scalar frame
\eqref{eq:constant-scalar-frame}, the Einstein and string metrics coincide,
so no constant dilaton factor appears.

We fix the chirality convention in the real orthonormal frame
\eqref{eq:real-orthonormal-frame}.  With $e_a,f_a$ the dual vectors, we set
\begin{equation}
  P=\Gamma_{e_1}\cdots\Gamma_{e_5}
    \Gamma_{f_1}\cdots\Gamma_{f_5},
  \qquad
  \Gamma_{11}=-P,
  \qquad
  \Gamma_{11}\epsilon=\epsilon.
  \label{eq:iib-chirality-convention}
\end{equation}
This is the positive Type IIB chirality convention used with the
orientation \eqref{eq:orientation} and the self-dual five-form.  The
complexified spinor doublet splits into the two eigenspaces of $\mathcal J$,
each of complex dimension sixteen:
\begin{equation}
  \mathcal K_{\gC}
  =
  \mathcal K_+\oplus\mathcal K_-,
  \qquad
  \mathcal J|_{\mathcal K_\pm}=\pm\ii,
  \qquad
  \dim_{\gC}\mathcal K_\pm=16.
  \label{eq:iib-doublet-eigenspaces}
\end{equation}
We first work on $\mathcal K_+$, where the two real Majorana--Weyl
parameters combine into one complex positive-chirality spinor.

The flux couples through the Clifford volume operators
\begin{equation}
  K_L
  =
  \Gamma_1\cdots\Gamma_5,
  \qquad
  K_R
  =
  \Gamma_{\bar1}\cdots\Gamma_{\bar5}.
  \label{eq:clifford-volumes}
\end{equation}
The Clifford relations give $K_L^2=-\ii$, $K_R^2=\ii$, and
$K_LK_R=\ii\Gamma_{11}$.  Raising each holomorphic flux index contributes
a factor $\ii$, while each antiholomorphic index contributes $-\ii$.
Consequently,
\begin{equation}
  \frac1{5!}F_{ABCDE}\Gamma^{ABCDE}
  =-\frac4L(K_L+K_R).
  \label{eq:holomorphic-flux-clifford}
\end{equation}
Substituting in \eqref{eq:gravitino-operator} on $\mathcal K_+$, where
$\mathcal J\epsilon=\ii\epsilon$, gives
\begin{equation*}
  \mathcal D_M\epsilon
  =\nabla_M\epsilon-\frac{\ii}{4L}(K_L+K_R)\Gamma_M\epsilon.
\end{equation*}
The volume relations imply $K_R=-K_L\Gamma_{11}$.  Since
$\Gamma_M\epsilon$ has negative chirality, we obtain
\begin{equation*}
  (K_L+K_R)\Gamma_M\epsilon
  =2K_L\Gamma_M\epsilon
  =2K_R\Gamma_M\epsilon.
\end{equation*}
We choose the expression with $K_L$ in the holomorphic block and the
equivalent expression with $K_R$ in the antiholomorphic block.  On these
positive-chirality spinors, the two derivatives are
\begin{align}
  \mathcal D_a
  &=
  \nabla_a
  -\frac{\ii}{2L}K_L\Gamma_a,
  \label{eq:left-killing-spinor-operator}
  \\
  \mathcal D_{\bar a}
  &=
  \nabla_{\bar a}
  -\frac{\ii}{2L}K_R\Gamma_{\bar a}.
  \label{eq:right-killing-spinor-operator}
\end{align}
Here $\mathcal D_a=\mathcal D_{E_a}$ and
$\mathcal D_{\bar a}=\mathcal D_{\bar E_a}$, with the same convention for
$\nabla$.  In this noncoordinate frame, the curvature includes the frame
commutator:
\begin{equation}
  \mathcal R^{\mathcal D}(U,V)
  =[\mathcal D_U,\mathcal D_V]-\mathcal D_{[U,V]}.
  \label{eq:supercovariant-curvature-definition}
\end{equation}
Using $K_L^2=-\ii$ and $K_R^2=\ii$, the supercovariant curvature vanishes
block by block.  With $\Gamma_{ab}=\tfrac12[\Gamma_a,\Gamma_b]$ and the
corresponding barred convention, the spin and flux contributions are
\begin{align}
  \mathcal R^{\mathcal D}(E_a,E_b)
  &=
  -\frac{\ii}{2L^2}\Gamma_{ab}
  +\frac{\ii}{2L^2}\Gamma_{ab}
  =0,
  \nonumber\\
  \mathcal R^{\mathcal D}(\bar E_a,\bar E_b)
  &=
  \frac{\ii}{2L^2}\Gamma_{\bar a\bar b}
  -\frac{\ii}{2L^2}\Gamma_{\bar a\bar b}
  =0,
  \nonumber\\
  \mathcal R^{\mathcal D}(E_a,\bar E_b)
  &=0.
  \label{eq:supercovariant-flatness}
\end{align}
The restriction of $\mathcal D$ to $\mathcal K_+$ is therefore flat and has sixteen
complex local parallel spinors.  Conjugation under the real structure
transports this result to $\mathcal K_-$.

\subsection{Global real Killing spinors}

We now impose the $\mathrm{IIB}_5$ reality condition on the Killing
spinors.  We denote the conjugation by
$c\psi=\mathcal B^{-1}_{(5,5)}\psi^*$.  In the real Clifford basis of
Appendix~\ref{app:clifford-check}, the spin connection and five-form terms
in $\mathcal D$ are real, since $F_5$ and $\mathcal J$ are real.
Consequently, $c\mathcal D=\mathcal D c$, and conjugation preserves the
Killing-spinor equation and chirality.  Since $\mathcal J$ has eigenvalues
$\pm\ii$, conjugation exchanges the two eigenspaces:
\begin{equation}
  c(\mathcal K_+)=\mathcal K_-,
  \qquad
  c(\mathcal K_-)=\mathcal K_+.
  \label{eq:reality-exchanges-j-eigenspaces}
\end{equation}
For each complex Killing spinor $\psi\in\mathcal K_+$, its conjugate
$c\psi$ therefore belongs to $\mathcal K_-$ and satisfies the same equation.
We obtain a real Killing spinor by taking
\begin{equation}
  \epsilon=\psi+c\psi.
  \label{eq:real-parallel-spinors}
\end{equation}
Every real solution has this form.  The sixteen complex integration
constants in $\psi$ give thirty-two independent real solutions.

These solutions extend globally.  The manifold $X_{\gR}\simeq T^*S^5$
retracts onto $S^5$, so it is simply connected and
$H^2(X_{\gR};\mathbb Z_2)=0$.  It is therefore spin
\cite{LawsonMichelsohn1989SpinGeometry}, and $c$ is defined on its global
spinor bundle.  The supercovariant connection is flat on both $\mathcal K_+$
and $\mathcal K_-$ and preserves the reality condition.  Parallel
transport therefore extends any real initial spinor uniquely over the
manifold.  We conclude that the background has thirty-two global Killing
spinors and is maximally supersymmetric.

\section{Killing superalgebra and supercoset}
\label{sec:superisometry}

\subsection{The real Killing superalgebra}

We now identify the algebra generated by the Killing vectors and the
thirty-two global real Killing spinors.  The complexified Killing
superalgebra is $\psl(4|4,\gC)$.  Conjugation exchanges the holomorphic
and antiholomorphic sectors and acts on the matrix blocks as
\begin{equation}
  \tau
  \begin{pmatrix}
    A&B\\ C&D
  \end{pmatrix}
  =
  \begin{pmatrix}
    \bar D&\bar C\\ \bar B&\bar A
  \end{pmatrix}.
  \label{eq:exchange-involution}
\end{equation}
Its fixed elements can be represented as
\begin{equation}
  \mathcal X(A,B)
  =
  \begin{pmatrix}
    A&B\\ \bar B&\bar A
  \end{pmatrix}
  \label{eq:exchange-fixed-matrix}
\end{equation}
modulo the projective central element.  The even generators have the form
$\operatorname{diag}(A,\bar A)$, with $A\in\slalg(4,\gC)$, and span a
thirty-dimensional real Lie algebra.  The odd matrix $B$ is an
unconstrained complex $4\times4$ matrix.  The real dimension is thus
\begin{equation}
  \dim_{\gR}\psl(4|4,\gC)^\tau
  =(30|32).
  \label{eq:exchange-dimension}
\end{equation}

The brackets commute with this involution, so the real Killing
superalgebra is
\begin{equation}
  \mathfrak g_{\tau}
  =
  \psl(4|4,\gC)^\tau.
  \label{eq:exchange-superalgebra}
\end{equation}
The even algebra acts transitively on the bosonic body, with stabilizer
$\spalg(4,\gC)$.  To see explicitly how the odd generators close, take two
odd elements $\mathcal X(0,B_1)$ and $\mathcal X(0,B_2)$.
Direct block multiplication gives the
real-bilinear bracket
\begin{equation}
  \left\{
    \mathcal X(0,B_1),
    \mathcal X(0,B_2)
  \right\}
  =
  \mathcal X
  \left(
    [B_1\bar B_2+B_2\bar B_1]_0,0
  \right),
  \label{eq:exchange-odd-bracket}
\end{equation}
where the subscript removes the projective trace.  Upon complexification,
the conjugate blocks become independent again, and
\eqref{eq:exchange-superalgebra} returns the full
$\psl(4|4,\gC)$ algebra.

To construct the supercoset, we specify the isotropy action.
The real stabilizer $H=\mathrm{Sp}(4,\gC)$ embeds in the
two complex factors as $h\mapsto(h,\bar h)$ and acts by
\begin{equation}
  B\longmapsto hB\bar h^{-1},
  \qquad
  H_{\gC}\simeq
  \mathrm{Sp}(4,\gC)_L\times\mathrm{Sp}(4,\gC)_R.
  \label{eq:real-isotropy-action}
\end{equation}
Appendix~\ref{app:full-indexed-algebra} gives the indexed brackets and their
normalization.

\subsection{The supercoset}

The symmetry supergroup $\mathcal G_{\tau}$, with even subgroup
$\mathrm{SL}(4,\gC)$, gives the superspace
\begin{equation}
  \mathcal M_{\tau}
  =\frac{\mathcal G_{\tau}}{\mathrm{Sp}(4,\gC)},
  \qquad
  \dim_{\gR}\mathcal M_{\tau}=(10|32).
  \label{eq:exchange-supercoset}
\end{equation}
Here we choose the connected symmetry group, with the stabilizer embedded
as in \eqref{eq:real-isotropy-action}.  The action on the odd matrix is
$B\mapsto gB\bar g^{-1}$.  The central element $-\mathbf1$ acts trivially
on both the bosonic coordinates and the odd generators.  Removing this
common $\mathbb Z_2$ from the symmetry group and stabilizer gives
\begin{equation}
  \mathcal G_{\tau}^{\mathrm{eff}}
  =\mathcal G_{\tau}/\{\pm\mathbf1\},
  \qquad
  \mathcal M_{\tau}
  \simeq\frac{\mathcal G_{\tau}^{\mathrm{eff}}}
       {\mathrm{Sp}(4,\gC)/\{\pm\mathbf1\}}.
  \label{eq:effective-exchange-supergroup}
\end{equation}
Its bosonic body may be written equivalently as
\begin{equation}
  \frac{\mathrm{SO}(6,\gC)}{\mathrm{SO}(5,\gC)}
  \simeq
  \frac{\mathrm{SL}(4,\gC)}{\mathrm{Sp}(4,\gC)}.
  \label{eq:effective-bosonic-body}
\end{equation}
The quotient thus has the realified quadric as its bosonic body and the
thirty-two odd directions generated by its global Killing spinors.

\section{Discussion and prospects}
\label{sec:comparison}

We have constructed a real, homogeneous solution of the untwisted
$\mathrm{IIB}_5$ theory
\cite{Hull1998Signature,GallMohaupt2021ArbitrarySignature} and determined
its thirty-two global Killing spinors and Killing superalgebra.  The
five-form supports a scalar-flat geometry with nonzero Ricci tensor.
Within the homogeneous ansatz, the field equations fix the phase of the
invariant tensors.

The realification of the complex sphere carries the anti-K\"ahler Einstein
metric $q+\bar q$ \cite{BorowiecEtAl1999AlmostComplex}.  Its phase-rotated
partner $\ii(\bar q-q)$ has the same connection but is not Einstein; the
five-form stress tensor supports the latter metric.  The same
underlying manifold $T^*S^5$ also admits the complete, positive-definite
Ricci-flat K\"ahler metric of Stenzel
\cite{Stenzel1993RicciFlat,ConlonHein2013ACCalabiYau}.
That metric is $\mathrm{SO}(6)$-invariant and Hermitian.  It lies outside
the full $\mathrm{SL}(4,\gC)$-invariant ansatz used here.

The complexified five-form is a sum of two orthogonal decomposable forms,
$\Omega$ and $\bar\Omega$, and satisfies the Pl\"ucker-type relations used
in the Lorentzian classification of maximally supersymmetric Type IIB
solutions
\cite{FigueroaOFarrillPapadopoulos2003Maximal,
FigueroaOFarrillPapadopoulos2004Pluecker}.  Reality exchanges their
five-dimensional distributions.  This gives a real form of the complex
Freund--Rubin geometry different from the product geometries obtained as
near-horizon limits of split-signature D3-branes
\cite{HullKhuri1999Worldvolume}.  Our analysis is restricted to this
complex Freund--Rubin branch and does not classify all maximally
supersymmetric solutions in split signature.

It would be interesting to construct a Green--Schwarz sigma model on the
real supercoset \cite{MetsaevTseytlin1998AdS5S5} and a pure-spinor
description extending the construction for $\AdS_5\times S^5$
\cite{Berkovits2000Covariant,BerkovitsChandia2001AdSVertices,
Vallilo2002OneLoop}.  The complex $\mathbb Z_4$ structure inherited from
the usual $\AdS_5\times S^5$ supercoset provides a starting point.  The
question is whether the kinetic and Wess--Zumino terms restrict to a real,
kappa-symmetric Green--Schwarz action on the fixed locus.  In the
pure-spinor formulation, the corresponding question concerns the reality
conditions on the ghost sector and their compatibility with the BRST
operator.

The split signature raises a further question about the physical string
spectrum.  In the local flat-space limit, the conventional light-cone
construction removes one spacelike and one timelike longitudinal
direction, leaving a transverse space of signature $(4,4)$.  With the
usual oscillator inner product, the remaining timelike excitations give
negative-norm states even after imposing the physical-state conditions.
We therefore expect additional negative-norm states in the string
spectrum on the present background.  A pure-spinor treatment
would allow this question to be studied through the BRST cohomology and
its inner product; this spectral analysis remains to be carried out.

The relation to complexified $\AdS_5\times S^5$ suggests investigating a
holographic description involving complexified $\mathcal N=4$
super-Yang--Mills theory, possibly on a real spacetime of signature
$(2,2)$.  The conventional SYM real form in this signature has noncompact
R-symmetry $\mathrm{SO}(3,3)$, with scalar kinetic signature $(3,3)$
\cite{HullKhuri1999Worldvolume}.  The two transverse gauge-field
polarizations have signature $(1,1)$, so the free bosonic modes have
kinetic signature $(4,4)$ per gauge-algebra generator, matching the
transverse string signature.  This also follows by reducing
ten-dimensional $(5,5)$ SYM along $(3,3)$ directions.  The modes with negative kinetic terms
could provide a boundary counterpart of the additional negative-norm
string states.  This agreement is suggestive, although it is so far
limited to counting free bosonic modes.

The independent conformal and R-symmetry algebras of this SYM real form
nevertheless differ from the factor-exchanging real
form found here.  Writing our even algebra as
$\mathfrak{sl}(4,\gR)\oplus\ii\mathfrak{sl}(4,\gR)$ suggests
associating these transformations with its real and imaginary parts, but
this is only a vector-space decomposition: the imaginary generators
close into the real ones, and the two parts do not commute.  A
prospective SYM description would therefore require a reality condition
intertwining conformal and internal transformations.  Identifying such
a condition, together with an appropriate boundary contour, remains a
speculative open problem.

\subsection*{Acknowledgements}
This work was partially supported by FONDECYT grant number 1250672. The author used OpenAI's Codex for scientific discussions, assisted
calculations, manuscript organization, and critical review. The author
takes responsibility for the results and conclusions presented here.

\appendix

\section{RR normalization dictionary}
\label{app:rr-conventions}

We record the metric variation and the conversion between the RR
normalizations used in the literature.  Varying the pseudo-action
\eqref{eq:iib5-pseudo-action} with respect to the inverse metric gives
\begin{align}
  R_{MN}-\frac12G_{MN}R
  &={}
  \frac1{4\cdot4!}
  F_{MPQRS}F_N{}^{PQRS}
  \nonumber\\*
  &\quad
  -\frac1{8\cdot5!}G_{MN}
  F_{PQRST}F^{PQRST}.
  \label{eq:pseudo-action-einstein-equation}
\end{align}
Self-duality makes $F_{PQRST}F^{PQRST}=0$, yielding the coefficient
$1/96$ in the Einstein equation used in the main text.

For comparison, the conventions of
\cite{DHokerGutperleKarchUhlemann2016WarpedAdS6,
DHokerGutperleUhlemann2025ComplexIIB} give
\begin{equation}
  R_{MN}
  =
  \frac16
  \widehat F_{MPQRS}\widehat F_N{}^{PQRS},
  \qquad
  \delta\psi_M
  =
  \nabla_M\epsilon
  -\frac1{480}
  \widehat F_{ABCDE}\Gamma^{ABCDE}\Gamma_M s^2\epsilon.
  \label{eq:dhoker-five-form-convention}
\end{equation}
Here $s^2=\begin{psmallmatrix}0&1\\-1&0\end{psmallmatrix}$ acts on the
spinor doublet.
Their recent comparison with \cite{BergshoeffEtAl2007AlternateRealities}
also notes the factor of four in the five-form coefficient.  In the present
truncation the complete dictionary is
\begin{equation}
  F_5=4\widehat F_5,
  \qquad
  s^2\longmapsto-\mathcal J.
  \label{eq:rr-normalization-dictionary}
\end{equation}
The second relation is the real change of doublet basis that reverses the
sign of one Majorana--Weyl component.  This change of basis preserves the
reality condition \eqref{eq:iib5-spinor-reality}.  Substitution gives
\begin{align}
  \frac16\widehat F_{MPQRS}\widehat F_N{}^{PQRS}
  &=
  \frac1{96}F_{MPQRS}F_N{}^{PQRS},
  \nonumber\\
  -\frac1{480}\widehat F_{ABCDE}\Gamma^{ABCDE}\Gamma_M s^2
  &=
  \frac1{16\cdot5!}F_{ABCDE}\Gamma^{ABCDE}\Gamma_M\mathcal J.
  \label{eq:rr-dictionary-check}
\end{align}
Since the three-form fields vanish here, the potential-level relation is
simply $C_4=4\widehat C_4$.  Multiplication by a real constant preserves the
reality assignment of $C_4$, so both notations describe the same untwisted
$\mathrm{IIB}_5$ branch.

\section{A real \texorpdfstring{$\mathrm{Cl}(5,5)$}{Cl(5,5)} normalization check}
\label{app:clifford-check}

We give a real Clifford representation that checks the phases in the
Killing-spinor calculation.  Let
\begin{equation}
  X=
  \begin{pmatrix}0&1\\1&0\end{pmatrix},
  \qquad
  Z=
  \begin{pmatrix}1&0\\0&-1\end{pmatrix},
  \qquad
  C=
  \begin{pmatrix}0&1\\-1&0\end{pmatrix}.
  \label{eq:real-pauli}
\end{equation}
Let $e_a,f_a$ be dual to the real coframe
\eqref{eq:real-orthonormal-frame}.  The following real $32\times32$ matrices generate $\mathrm{Cl}(5,5)$:
\begin{align}
  \Gamma_{e_a}
  &=
  Z^{\otimes(a-1)}
  \otimes X
  \otimes\mathbf1^{\otimes(5-a)},
  \nonumber\\
  \Gamma_{f_a}
  &=
  Z^{\otimes(a-1)}
  \otimes C
  \otimes\mathbf1^{\otimes(5-a)},
  \qquad
  a=1,\ldots,5.
  \label{eq:cl55-representation}
\end{align}
They satisfy
\begin{equation}
  \Gamma_{e_a}^2=1,
  \qquad
  \Gamma_{f_a}^2=-1,
  \qquad
  \{\Gamma_A,\Gamma_B\}=0
  \quad(A\neq B).
  \label{eq:cl55-relations}
\end{equation}
We define the spinor complex-conjugation matrix $\mathcal B_{(5,5)}$ by
\begin{equation}
  (\Gamma_A)^*
  =\mathcal B_{(5,5)}\Gamma_A\mathcal B_{(5,5)}^{-1},
  \qquad
  \mathcal B_{(5,5)}^*\mathcal B_{(5,5)}=\mathbf1_{32},
  \label{eq:cl55-conjugation-definition}
\end{equation}
where $A$ labels the real frame vectors $e_a,f_a$ and $*$ denotes
entrywise complex conjugation.  All three matrices $X,Z,C$ are real,
so their tensor products in \eqref{eq:cl55-representation} obey
$(\Gamma_A)^*=\Gamma_A$.  The identity therefore satisfies both defining
relations, and we choose
\begin{equation}
  \mathcal B_{(5,5)}=\mathbf1_{32},
  \qquad
  \mathcal B_{(5,5)}^{-1}\epsilon^*=\epsilon^*=\epsilon.
  \label{eq:cl55-conjugation-identity}
\end{equation}
The chirality operator \eqref{eq:iib-chirality-convention} is also real,
so this condition preserves each Weyl module.  On the positive-chirality
module, $\mathcal B_{(5,5)}$ restricts to $\mathbf1_{16}$ and imposes
componentwise reality on each member of the Type IIB doublet.

The matrices associated with the vectors dual to $E^a,\bar E^a$ are
\begin{equation}
  \Gamma_a
  =
  \frac{1-\ii}{2}\Gamma_{e_a}
  +\frac{1+\ii}{2}\Gamma_{f_a},
  \qquad
  \Gamma_{\bar a}
  =
  \frac{1+\ii}{2}\Gamma_{e_a}
  +\frac{1-\ii}{2}\Gamma_{f_a}.
  \label{eq:complex-gamma}
\end{equation}
The real coframe gives the orientation explicitly,
\begin{equation}
  \ii\Omega\wedge\bar\Omega
  =-e^1\wedge\cdots\wedge e^5\wedge f^1\wedge\cdots\wedge f^5.
  \label{eq:real-frame-orientation}
\end{equation}
Direct multiplication in this representation gives
\begin{equation}
  P^2=1,
  \qquad
  K_L^2=-\ii,\qquad K_R^2=\ii,
  \qquad
  K_LK_R=-\ii P=\ii\Gamma_{11}.
  \label{eq:cl55-chirality-check}
\end{equation}
Thus the positive-chirality module defined in
\eqref{eq:iib-chirality-convention} is the $P=-1$ eigenspace, with projector
$\Pi_+=(1+\Gamma_{11})/2=(1-P)/2$ and real dimension sixteen.  Substituting
the flux \eqref{eq:complex-frame-background} into
\eqref{eq:gravitino-operator} reproduces
\eqref{eq:left-killing-spinor-operator} and
\eqref{eq:right-killing-spinor-operator} on this eigenspace.  Matrix
multiplication then gives
\begin{equation}
  \mathcal R^{\mathcal D}(E_a,E_b)\Pi_+
  =\mathcal R^{\mathcal D}(\bar E_a,\bar E_b)\Pi_+
  =\mathcal R^{\mathcal D}(E_a,\bar E_b)\Pi_+=0.
  \label{eq:cl55-projected-flatness}
\end{equation}
\section{The full indexed \texorpdfstring{$\psl(4|4,\gC)$}{psl(4|4,C)} algebra}
\label{app:full-indexed-algebra}

The block bracket \eqref{eq:exchange-odd-bracket} determines the
factor-exchanging real form, but for component calculations it is useful to
display its complexification in an $\mathrm{SO}(5,\gC)$ basis.  We use the
even generators
\begin{equation}
  J_{ab},\quad T_a,
  \qquad
  \bar J_{ab},\quad \bar T_a,
  \qquad
  a,b=1,\ldots,5,
  \label{eq:app-full-even-generators}
\end{equation}
with $J_{ab}=-J_{ba}$ and $\bar J_{ab}=-\bar J_{ba}$.  Before the
factor-exchanging reality condition is imposed, the bar labels the second
complex block and does not denote coefficient conjugation.

Each complex odd sector transforms as $(\mathbf4,\mathbf4)$ under the
two factors of the complexified isotropy in
\eqref{eq:real-isotropy-action}, using the symplectic forms to identify the
dual representations.  For component calculations, we identify the two
symplectic bases and hence their vector representations.  Under the
auxiliary diagonal subgroup $\mathrm{Sp}(4,\gC)_\Delta$, we have
\begin{equation}
  \mathbf4\otimes\mathbf4
  =
  \mathbf1\oplus\mathbf5\oplus\mathbf{10}.
  \label{eq:odd-sp4-branching}
\end{equation}
The $\mathbf5$ and $\mathbf{10}$ here label odd components, with the same
representation types as the even transvections and rotations.  This
branching organizes the indexed algebra; it is not a decomposition under
the full real isotropy.
We raise and lower the common vector indices with
$\delta_{ab}$ and normalize $\epsilon_{12345}=1$.

The two sixteen-dimensional odd sectors of the complex $\mathbb Z_4$
grading decompose under this auxiliary diagonal subgroup as
\begin{equation}
  \begin{array}{c|ccc}
    &\mathbf1&\mathbf5&\mathbf{10}\\ \hline
    \mathfrak g_1
      &Q&Q_a&Q_{ab}\\
    \mathfrak g_3
      &\bar Q&\bar Q_a&\bar Q_{ab}
  \end{array},
  \qquad
  Q_{ab}=-Q_{ba},
  \qquad
  \bar Q_{ab}=-\bar Q_{ba}.
  \label{eq:app-full-odd-generators}
\end{equation}
Together these components account for all thirty-two complex odd
generators of the complex superalgebra.  The factor-exchanging involution relates
the two complex blocks and selects a thirty-two-real-dimensional fixed
subspace.

We choose the normalization in which the two bosonic algebras are
\begin{align}
  [J_{ab},J_{cd}]
  &={}
  \delta_{bc}J_{ad}
  -\delta_{bd}J_{ac}
  -\delta_{ac}J_{bd}
  +\delta_{ad}J_{bc},
  \nonumber\\
  [J_{ab},T_c]
  &={}
  \delta_{bc}T_a-\delta_{ac}T_b,
  &
  [T_a,T_b]
  &={}
  J_{ab},
  \label{eq:app-full-left-even}\\
  [\bar J_{ab},\bar J_{cd}]
  &={}
  \delta_{bc}\bar J_{ad}
  -\delta_{bd}\bar J_{ac}
  -\delta_{ac}\bar J_{bd}
  +\delta_{ad}\bar J_{bc},
  \nonumber\\
  [\bar J_{ab},\bar T_c]
  &={}
  \delta_{bc}\bar T_a-\delta_{ac}\bar T_b,
  &
  [\bar T_a,\bar T_b]
  &={}
  \bar J_{ab}.
  \label{eq:app-full-right-even}
\end{align}
All brackets between the barred and unbarred even sectors vanish.

The rotations act on the first odd sector according to
\begin{align}
  [J_{ab},Q]
  &={}
  \frac12Q_{ab},
  &
  [\bar J_{ab},Q]
  &={}
  -\frac12Q_{ab},
  \label{eq:app-full-rotation-scalar}\\
  [J_{ab},Q_c]
  &={}
  \frac12
  \left(
    \delta_{bc}Q_a-\delta_{ac}Q_b
  \right)
  -\frac14\epsilon_{abcde}Q_{de},
  \nonumber\\
  [\bar J_{ab},Q_c]
  &={}
  \frac12
  \left(
    \delta_{bc}Q_a-\delta_{ac}Q_b
  \right)
  +\frac14\epsilon_{abcde}Q_{de},
  \label{eq:app-full-rotation-vector}\\
  [J_{ab},Q_{cd}]
  &={}
  \frac12
  \left(
    \delta_{bc}Q_{ad}
    -\delta_{bd}Q_{ac}
    -\delta_{ac}Q_{bd}
    +\delta_{ad}Q_{bc}
  \right)
  \nonumber\\
  &\quad
  +\frac12\epsilon_{abcde}Q_e
  +\frac12
  \left(
    \delta_{ad}\delta_{bc}
    -\delta_{ac}\delta_{bd}
  \right)Q,
  \nonumber\\
  [\bar J_{ab},Q_{cd}]
  &={}
  \frac12
  \left(
    \delta_{bc}Q_{ad}
    -\delta_{bd}Q_{ac}
    -\delta_{ac}Q_{bd}
    +\delta_{ad}Q_{bc}
  \right)
  \nonumber\\
  &\quad
  -\frac12\epsilon_{abcde}Q_e
  -\frac12
  \left(
    \delta_{ad}\delta_{bc}
    -\delta_{ac}\delta_{bd}
  \right)Q.
  \label{eq:app-full-rotation-bivector}
\end{align}
The same relations hold after the simultaneous replacement
\begin{equation}
  (Q,Q_a,Q_{ab})
  \longleftrightarrow
  (\bar Q,\bar Q_a,\bar Q_{ab}).
  \label{eq:app-full-grade-replacement}
\end{equation}

The transvections exchange the two odd sectors.  Their nonzero action on
the first one is
\begin{align}
  [T_a,Q]
  &={}
  \frac12\bar Q_a,
  &
  [\bar T_a,Q]
  &={}
  -\frac12\bar Q_a,
  \label{eq:app-full-transvection-scalar}\\
  [T_a,Q_b]
  &={}
  \frac12\delta_{ab}\bar Q
  +\frac12\bar Q_{ab},
  &
  [\bar T_a,Q_b]
  &={}
  -\frac12\delta_{ab}\bar Q
  +\frac12\bar Q_{ab},
  \label{eq:app-full-transvection-vector}\\
  [T_a,Q_{bc}]
  &={}
  \frac12
  \left(
    \delta_{ab}\bar Q_c
    -\delta_{ac}\bar Q_b
  \right)
  -\frac14\epsilon_{abcde}\bar Q_{de},
  \nonumber\\
  [\bar T_a,Q_{bc}]
  &={}
  \frac12
  \left(
    \delta_{ab}\bar Q_c
    -\delta_{ac}\bar Q_b
  \right)
  +\frac14\epsilon_{abcde}\bar Q_{de}.
  \label{eq:app-full-transvection-bivector}
\end{align}
These equations also hold with the two triples in
\eqref{eq:app-full-grade-replacement} interchanged.

The anticommutators within each odd sector are
\begin{align}
  \{Q,Q\}
  &={}
  \{Q,Q_{ab}\}
  =
  \{Q_a,Q_b\}
  =0,
  \nonumber\\
  \{Q,Q_a\}
  &={}
  -2(T_a+\bar T_a),
  \nonumber\\
  \{Q_a,Q_{bc}\}
  &={}
  2
  \left[
    \delta_{ab}(T_c-\bar T_c)
    -\delta_{ac}(T_b-\bar T_b)
  \right],
  \nonumber\\
  \{Q_{ab},Q_{cd}\}
  &={}
  2\epsilon_{abcde}(T_e+\bar T_e),
  \label{eq:app-full-first-grade-anticommutators}\\
  \{\bar Q,\bar Q\}
  &={}
  \{\bar Q,\bar Q_{ab}\}
  =
  \{\bar Q_a,\bar Q_b\}
  =0,
  \nonumber\\
  \{\bar Q,\bar Q_a\}
  &={}
  2(T_a+\bar T_a),
  \nonumber\\
  \{\bar Q_a,\bar Q_{bc}\}
  &={}
  -2
  \left[
    \delta_{ab}(T_c-\bar T_c)
    -\delta_{ac}(T_b-\bar T_b)
  \right],
  \nonumber\\
  \{\bar Q_{ab},\bar Q_{cd}\}
  &={}
  -2\epsilon_{abcde}(T_e+\bar T_e).
  \label{eq:app-full-third-grade-anticommutators}
\end{align}

Finally, the mixed anticommutators are
\begin{align}
  \{Q,\bar Q\}
  &={}
  \{Q,\bar Q_a\}
  =
  \{Q_a,\bar Q\}
  =0,
  \nonumber\\
  \{Q,\bar Q_{ab}\}
  &={}
  -2(J_{ab}+\bar J_{ab}),
  &
  \{Q_{ab},\bar Q\}
  &={}
  2(J_{ab}+\bar J_{ab}),
  \nonumber\\
  \{Q_a,\bar Q_b\}
  &={}
  2(J_{ab}-\bar J_{ab}),
  \label{eq:app-full-mixed-low}\\
  \{Q_a,\bar Q_{bc}\}
  &={}
  \epsilon_{abcde}(J_{de}+\bar J_{de}),
  \nonumber\\
  \{Q_{ab},\bar Q_c\}
  &={}
  -\epsilon_{abcde}(J_{de}+\bar J_{de}),
  \nonumber\\
  \{Q_{ab},\bar Q_{cd}\}
  &={}
  -2
  \left[
    \delta_{bc}(J_{ad}-\bar J_{ad})
    -\delta_{bd}(J_{ac}-\bar J_{ac})
  \right.
  \nonumber\\
  &\qquad\left.
    -\delta_{ac}(J_{bd}-\bar J_{bd})
    +\delta_{ad}(J_{bc}-\bar J_{bc})
  \right].
  \label{eq:app-full-mixed-high}
\end{align}

Equations~\eqref{eq:app-full-left-even}--
\eqref{eq:app-full-mixed-high} give the complete complex
$\psl(4|4,\gC)$ bracket in this basis.  They realize the usual
$\mathbb Z_4$ pattern: rotations close on rotations, transvections square
to rotations, two odd generators give an even generator, and a
transvection exchanges the odd sectors.  The factor-exchanging real form of
Section~\ref{sec:superisometry} is the fixed set of the matrix involution
\eqref{eq:exchange-involution}; it changes the reality condition on this
algebra, not its complex structure constants.

\bibliographystyle{abe}
\bibliography{references}

\end{document}